\documentclass[12pt,a4paper]{article}

\usepackage{amsmath}
\usepackage{amssymb}
\usepackage{hyperref}
\usepackage{hyperref}
\usepackage{graphics}
\usepackage{graphicx}
\usepackage{epsfig}

\let\oldappendix=\appendix
\let\oldsection=\section
\renewcommand{\appendix}{\oldappendix%
\def\theequation{\Alph{section}.\arabic{equation}}%
\renewcommand{\section}{\setcounter{equation}{0}\oldsection}}

\newcommand{\beq}{\begin{equation}}
\newcommand{\eeq}{\end{equation}}
\newcommand{\beqa}{\begin{eqnarray}}
\newcommand{\eeqa}{\end{eqnarray}}
\newcommand{\no}{\nonumber}
\newcommand{\q}{\quad}
\newcommand{\qq}{\qquad}
\newcommand{\mnod}{\stackrel{\circ}{M}}

\begin{document}

\hfill 

\hfill 

\bigskip\bigskip

\begin{center}

{{\Large\bf  Final state interactions in nonleptonic hyperon decays}}

\end{center}

\vspace{.4in}

\begin{center}
{\large B. Borasoy\footnote{email: borasoy@physik.tu-muenchen.de},
 E. Marco}

\bigskip

\bigskip

\href{http://www.ph.tum.de/}{Physik Department}\\
\href{http://www.tum.de/}{Technische Universit{\"a}t M{\"u}nchen}\\
D-85747 Garching, Germany \\

\vspace{.2in}

\end{center}

\vspace{.7in}

\thispagestyle{empty} 

\begin{abstract}
We investigate the importance of final state interactions in weak nonleptonic
hyperon decays within a relativistic chiral unitary
approach based on coupled channels. The effective potentials for meson-baryon 
scattering are derived from a chiral effective Lagrangian and iterated 
in a Bethe-Salpeter equation, which generates the low lying baryon resonances
dynamically. The inclusion of final state
interactions decreases the discrepancy between theory and experiment for
both $s$ and $p$ waves.
Our study indicates that contributions from higher order
terms of the weak effective Lagrangian
may play an important role in these decays. 

\end{abstract}\bigskip

\begin{center}
\begin{tabular}{ll}
\textbf{PACS:}&12.39.Fe, 13.30.-a, 14.20.Jn\\[6pt]
\textbf{Keywords:}& nonleptonic hyperon decays, chiral symmetry,
 unitarity, resonances.
\end{tabular}
\end{center}

%

\vfill

\section{Introduction}\label{sec:intro}
Nonleptonic hyperon decays have been a topic of interest for more than three
decades and a convincing theoretical description is still missing.
There are seven such transitons
and after omitting final state interactions their matrix elements are usually 
described in terms of two amplitudes---the
parity violating $s$ wave and the parity conserving $p$ wave.
A convenient framework to study these decays is provided by chiral perturbation
theory (ChPT) whereby
the amplitudes are expanded in terms of small four-momenta and the current
quark masses $m_q$ of the light quarks $q= u,d,s$.
At lowest order in this expansion the amplitudes are given in terms of 
two unknown coupling constants of the effective Lagrangian, so-called
low-energy constants (LECs). It is well-known that with just these two LECs
it is not possible to obtain a reasonable description of both $s$ and $p$
waves \cite{DGH}. If, {\it e.g.}, one employs values which provide a
reasonable fit to the $s$ waves, a poor description for the $p$ waves is
obtained. A good $p$ wave representation,
on the other hand, yields a poor $s$ wave fit.
Different authors have tried to overcome this problem by going beyond leading order.
In \cite{Bij} the leading chiral logarithms to these decays were calculated
neglecting local counterterms,
but the resulting $s$ wave predictions no longer agreed with the data,
and the corrections to the $p$ waves were even larger.

The decays were reinvestigated by Jenkins \cite{Jen} within the framework
of heavy baryon ChPT, explicitly taking into account spin-$3/2$ decuplet loops.
Again, no local counterterms were included and only the non-analytic pieces
of the meson loops were retained. She found significant cancellations between
the octet and decuplet components in the loops, restoring agreement
between theory and experiment for the $s$ waves, although for the $p$ waves
the chiral corrections did not provide a good description of the data.
Hence, the inability to fit $s$ and $p$ waves simultaneously still remained
even after including the lowest non-analytic contributions.

A complete calculation at the one-loop level and including local counterms was
performed in \cite{BH1}. Due to the proliferation of new unknown LECs at higher
chiral orders a fit to both $s$ and $p$ waves is possible, but not unique,
so that the theory lacks predictive power. However, it remains unclear, whether the
values of the coupling constants have been chosen appropriately.

Another intriguing possibility was examined by Le Yaouanc {\it et al.}, who
assert that a reasonable fit for both $s$ and $p$ waves can be provided by
adding pole contributions from $SU(6) (70, 1^-)$ states to the $s$ waves
\cite{LeY}. Their calculations were performed in a simple constituent quark
model and appear to be able to provide a resolution to the $s$ and $p$ wave
dilemma. This approach has been studied also within the framework of ChPT in
\cite{BH2}. To this end, the spin-$1/2^-$ octet from the $(70, 1^-)$ states
and the octet of Roper-like $1/2^+$ fields have been included in the effective
field theory. In ChPT the inclusion of the Roper octet which is in the same
mass range as the $1/2^-$ octet was necessary to improve the agreement with
the experimental data, since the two lowest order couplings for the $p$ wave
amplitudes tend to cancel  thus enhancing the contributions from terms of
higher chiral order \footnote{In \cite{LeY}, on the other hand, the inclusion
of the $(70, 1^-)$ states was sufficient to obtain a satisfactory fit to both
$s$ and $p$ waves, since in the quark model the expressions for the $p$ waves
include additional explicit $SU(3)$ symmetry breaking corrections of second
chiral order, in which case a much improved fit to the $p$ waves is possible.}.
Integrating out the resonances generates counterterms of the Lagrangian at
next-to-leading order and by fitting the weak couplings of the resonances
one obtains satisfactory agreement with experiment. Thus the inclusion of
spin-$1/2$ resonances in nonleptonic hyperon decays provides an  estimate
of the importance of higher order counterterms.

There is, however, an independent way of investigating the importance of
low lying resonances in nonleptonic hyperon decays whereby 
the resonances are produced dynamically within a relativistic chiral unitary
approach based on coupled channels. 
In this framework the hyperon decays via
the weak interaction into a Goldstone boson,
$(\pi, K, \eta)$, and a ground state baryon which can then undergo final
state interactions. These can be accurately described by deriving 
effective potentials for meson-baryon 
scattering from the chiral effective Lagrangian and subsequently 
iterating them in a Bethe-Salpeter equation (BSE).
The approach has been successfully applied both to meson-baryon scattering 
processes and photoproduction of pseudoscalar mesons
and the pertinent resonances have been observed \cite{KSW, KWW, OR, OM, BMW}.
In the purely mesonic sector, work along
these lines has been successfully applied, {\it e.g.},
to radiative $\phi$ decays \cite{Oller,MH}.
Here, we extend this method to the weak sector by appending to the weak hyperon decays 
at the tree level strong final state interactions, which accounts for the
exchange of resonances without including them explicitly. The chiral parameters
of the effective potentials are chosen in such a way that they reproduce
the experimental phase shifts of pion-nucleon scattering in the energy range
we are interested in. Once these parameters have been fixed, the only remaining
unknown LECs are the two weak couplings of the leading ground state
baryon Lagrangian. The resulting weak matrix elements acquire imaginary
components and can no longer be described in terms of just two but rather three
independent  amplitudes.
Our model can help to clarify the role of low lying baryon resonances for
nonleptonic hyperon decays and can answer the question, whether omission
of final state interactions was justified in previous work.

The work is organized as follows. In the next section, we introduce the
effective Lagrangian of the strong and weak interactions and derive the
tree level results.
The coupled channel method is explained in Sec.~3. In Sec.~4,
after having matched the chiral parameters of the strong interactions
to the experimental phase shifts of pion-nucleon scattering,
we implement final state interactions in nonleptonic hyperon decays
and discuss the role of resonances. Sec.~5 contains our summary and
conclusions.

\section{Nonleptonic hyperon decays} \label{sec:nonlep}

There exist seven experimentally accessible nonleptonic hyperon decays: 
$\Sigma^+  \rightarrow n \, \pi^+ \, , \, 
\Sigma^+  \rightarrow p \, \pi^0 \, , \, 
\Sigma^- \rightarrow n \, \pi^- \, , \, 
\Lambda \rightarrow p \, \pi^- \, ,  \,
\Lambda \rightarrow n \, \pi^0 \, , \,
\Xi^- \rightarrow  \Lambda \, \pi^-\,\mbox{and} \, 
\Xi^0 \rightarrow  \Lambda \, \pi^0 $,
and the matrix elements of these decays can each be expressed in terms of a
parity violating $s$ wave amplitude $ A_{ij}$
and a parity conserving $p$ wave amplitude $ B_{ij}$
\beq
{\cal A}( B_i \rightarrow B_j \, \pi) =
\bar{u}_{B_j} \Big\{ \, A_{ij} + \, B_{ij}\gamma_5 \Big\}u_{B_i}.
\eeq
The underlying strangeness-changing Hamiltonian
transforms under $SU(3) \times SU(3)$ as $(8_L, 1_R) \oplus (27_L,1_R)$
and, experimentally, the octet piece dominates over the 27-plet by a factor
of twenty or so. Consequently, we will neglect the 27-plet in what follows.
Isospin symmetry of the strong interactions implies then the relations 
\beqa \label{iso}
&&
{\cal A}(\Lambda \rightarrow p \, \pi^-)
+ \sqrt{2} \, {\cal A}(\Lambda \rightarrow n \, \pi^0) = 0 \no \\
&&
{\cal A}(\Xi^- \rightarrow  \Lambda \, \pi^-)
+ \sqrt{2} \, {\cal A}(\Xi^0 \rightarrow  \Lambda \, \pi^0) = 0 \no \\
&&
\sqrt{2} \, {\cal A}(\Sigma^+ \rightarrow p \, \pi^0)
+ {\cal A}(\Sigma^- \rightarrow n \, \pi^-)
- {\cal A}(\Sigma^+ \rightarrow n \, \pi^+) = 0  ,
\eeqa
which hold for both $s$ and $p$ waves.
We choose
$\Sigma^+  \rightarrow n \, \pi^+ \, , \, 
\Sigma^- \rightarrow n \, \pi^- \, , \, 
\Lambda \rightarrow p \, \pi^- \, \mbox{and} \,
\Xi^- \rightarrow  \Lambda \, \pi^- $
as the four independent decay amplitudes which are not related by isospin.
With our sign conventions one defines the amplitudes $s$ and $p$ \cite{pdg}
\beq
s = A , \qq
p = \frac{|{\bf p}_f|}{E_f + m_f} B 
\eeq
which are related to the decay width $\Gamma$ and the decay parameters
$\alpha, \beta, \gamma$ by
\beqa  \label{observ}
\Gamma &=& \frac{|{\bf p}_f| (E_f + m_f)}{4 \pi m_i} (|s|^2 + |p|^2) ,
\qq  \alpha = - \frac{2 \mbox{Re} (s^* p)}{|s|^2 + |p|^2} ,  \no \\
\beta &=& - \frac{2 \mbox{Im} (s^* p)}{|s|^2 + |p|^2} , \qq
\phi = \arcsin (\frac{\beta}{\sqrt{1 - \alpha^2}}) ,
\eeqa
where ${\bf p}_f, E_f, m_f$ are the three-momentum, energy and mass of the
final
baryon, respectively, and $m_i$ is the mass of the incoming baryon.
Only three of the four observables in Eq.~(\ref{observ}) are independent
and if one omits
final state interactions, the amplitudes $s$ and $p$ are
real, and $\beta$ and $\phi$ vanish.
Previous fits to nonleptonic hyperon decays have always been performed under
this assumption, and the imaginary components of the amplitudes were neglected.
In our case, since we include final state interactions,
it is more convenient to work directly with the real observables
$\Gamma, \alpha, \beta$ and $\gamma$ which are quoted in experiments
\cite{pdg}.

Our starting point is the relativistic effective chiral 
strong interaction Lagrangian for the 
pseudoscalar bosons coupled to the lowest--lying 1/2$^+$ baryon octet,
which reads at lowest chiral order
\beq
{\cal L}_{\phi B}^{(1)}  
 = i \langle \bar{B} \gamma_{\mu} [D^{\mu},B] \rangle 
 - \mnod \langle \bar{B}B \rangle 
- \frac{1}{2} D \langle \bar{B} \gamma_{\mu}
 \gamma_5 \{u^{\mu},B\} \rangle  
- \frac{1}{2} F \langle \bar{B} \gamma_{\mu} \gamma_5 [u^{\mu},B] \rangle ,
\eeq
where the superscript denotes the chiral order, $\mnod$ is the octet
baryon mass in the chiral limit
and $\langle \ldots \rangle$ denotes the trace in flavor space. 
The values of $D$ and $F$ are extracted phenomenologically
from the semileptonic hyperon decays and a fit to data delivers $D= 0.80 
\pm 0.01$, $F=0.46 \pm 0.01$ \cite{CR}.
The pseudoscalar Goldstone fields ($\phi = \pi, K, \eta$) are collected in
the  $3 \times 3$ unimodular, unitary matrix $U(x)$, 
\begin{equation}
 U(\phi) = u^2 (\phi) = \exp \lbrace 2 i \phi / f_\pi \rbrace  \qq ,\qq
u_{\mu} = i u^\dagger \nabla_{\mu} U u^\dagger
\end{equation}
where $f_\pi \simeq 92.4$ MeV is the pion decay constant,
\begin{equation}
 \phi =  \frac{1}{\sqrt{2}}  \left(\begin{array}{ccc}
 \frac{1}{ \sqrt 2} \pi^0 + \frac{1 }{ \sqrt 6} \eta &\pi^+ &K^+ \\
\pi^-& -\frac{1}{ \sqrt 2} \pi^0 + \frac{1}{ \sqrt 6} \eta & K^0\\
K^-   &  \bar{K^0}&- \frac{2}{\sqrt 6} \eta \end{array} \right)
 \end{equation}
represents the contraction of the pseudoscalar fields with the Gell-Mann
matrices and
$B$ is the standard $SU(3)$ matrix representation of the low--lying
spin--1/2 baryons $( N, \Lambda, \Sigma, \Xi)$.

The leading strong effective Lagrangian is in general not sufficient
to obtain an appropriate description of meson-baryon scattering and one
needs to go beyond leading order \cite{KWW, BMW}.
At next-to-leading order the terms relevant for meson-baryon
scattering are 
\begin{eqnarray}  \label{bar2}
{\cal L}_{\phi B}^{(2)} &=&  b_D \langle \bar{B}  \{\chi_+,B\} \rangle +
b_F \langle \bar{B}  [\chi_+,B] \rangle + b_0 \langle \bar{B}B \rangle
\langle \chi_+ \rangle \no \\[3pt]
&+& d_1 \langle \bar{B} \{v \cdot u,[v \cdot u,B]\}  \rangle 
+ d_2 \langle \bar{B} [v \cdot u,[v \cdot u,B]]  \rangle
+ d_3 \langle \bar{B} v \cdot u \rangle  \langle v \cdot u B  \rangle
+ d_4 \langle \bar{B} B \rangle  \langle (v \cdot u) ( v \cdot u)
 \rangle \no \\[3pt]
&+& g_1 \langle \bar{B} \{{\bf u},[{\bf u},B]\}  \rangle 
+ g_2 \langle \bar{B} [{\bf u},[{\bf u},B]]  \rangle
+ g_3 \langle \bar{B} {\bf u} \rangle  \langle {\bf u} B  \rangle
+ g_4 \langle \bar{B} B \rangle  \langle {\bf u}  {\bf u} \rangle \no \\[3pt]
&+& i h_1 \langle \bar{B} \sigma_{\mu \nu}\{[u^\mu,u^\nu],B\}  \rangle
+ i h_2 \langle \bar{B} \sigma_{\mu \nu}[[u^\mu,u^\nu],B]  \rangle
+ i h_3 \langle \bar{B} u^\mu \rangle  \sigma_{\mu \nu} \langle u^\nu B  \rangle  ,
\end{eqnarray}
where we made use of a Cayley-Hamilton identity, in order to eliminate 
$\langle \bar{B} \{v \cdot u,\{v \cdot u,B\}\}  \rangle $ and
$\langle \bar{B} \{{\bf u},\{{\bf u},B\}\}  \rangle $, and $v$ is a four-velocity
with $v^2=1$.
We work in the 
isospin limit $m_u = m_d = \hat{m}$ and
explicit chiral symmetry breaking is induced via $\chi_+ = u^\dagger \chi u^\dagger + 
u \chi^\dagger u$, with $\chi = \mbox{diag}(m_{\pi}^2, m_{\pi}^2,
2 m_{K}^2 - m_{\pi}^2)$. 
Instead of the counterterm structures $\langle  \bar{B} {\bf u} \cdot {\bf u} B \rangle $
and $\langle  \bar{B} (v \cdot u) ( v \cdot u) B \rangle $
one may employ $\langle  \bar{B} u_\mu u^\mu    B \rangle $,
$\langle  \bar{B} u^\mu u^\nu \gamma_\mu  [D_\nu,  B ]\rangle $
and $\langle  \bar{B} u^\mu u^\nu [D_\mu , [D_\nu,  B ]]\rangle $.
The contact terms of the strong effective Lagrangian, however, will be used to derive
the effective potentials of meson-baryon scattering for on-shell particles
in which case the latter two terms reduce to $\langle  \bar{B} u_0^2 B \rangle $
modulo higher order corrections which are beyond the accuracy of the present investigation.
One can thus choose to work with the two combinations
$(v \cdot u)^2 = u_0^2$ and ${\bf u} \cdot {\bf u}=  u_0^2 - u_\mu \cdot 
 u^\mu$ for the mesonic fields, see also
\cite{KWW}.

The LECs $b_D$ and $b_F$ are responsible for the splitting of the
baryon octet masses 
at leading order in symmetry breaking,
\begin{eqnarray} \label{massdiff}
M_\Sigma - M_N &=& 4 (b_D -b_F) (m_K^2 -m_\pi^2) \no \\
M_\Xi - M_N &=& - 8 b_F (m_K^2 -m_\pi^2) \no \\
M_\Sigma - M_\Lambda &=& \frac{16}{3} b_D (m_K^2 -m_\pi^2).
\end{eqnarray}
Since the three baryon mass differences are represented in terms of
two parameters,
there is a corresponding sum rule---the Gell-Mann--Okubo mass relation
for the baryon octet \cite{GMO}:
\begin{equation}
M_\Sigma - M_N = \frac{1}{2} (M_\Xi - M_N) + \frac{3}{4}
(M_\Sigma - M_\Lambda )
\end{equation}
which experimentally has only a 3\% deviation.
A least-squares fit to
the mass differences (\ref{massdiff}) yields
$b_D = 0.066 $ GeV$^{-1}$ and $b_F = -0.213 $ GeV$^{-1}$, but we will allow for
small variations of these values in the fitting procedure to the experimental
phase shifts, in order to account for higher order
contributions to the baryon masses.
The $b_0$ term, on the other hand, cannot be determined from the masses alone.
One needs further information which is provided by the pion-nucleon
$\sigma$-term and reads at leading order
\begin{equation}
\sigma_{\pi N} = \hat{m} \langle N |   \bar{u} u + \bar{d} d | N \rangle 
=  - 2 m_\pi^2 ( b_D +b_F + 2 b_0).
\end{equation}
Employing the empirical value of \cite{GLS}, $\sigma_{\pi N}= 45 \pm 8$ MeV,
one obtains $b_0= -0.52 \pm 0.10$ GeV$^{-1}$. Recently, this value has
been questioned \cite{PSWA} and the authors of this work arrive at a
value  $\sigma_{\pi N}= 60 \pm 7$ MeV which would translate into a
value of $b_0= -0.71 \pm 0.09 $ GeV$^{-1}$. In particular the
latter value yields a large strangeness content of the proton, however
both values may change if loop effects are included \cite{BM}. 
Due to these uncertainties in the value of $b_0$, any
fitted value which lies in the range $-0.80$ 
GeV$^{-1}< b_0 < -0.20$ GeV$^{-1} $ is still acceptable.
The situation is less clear for the derivative terms $d_i , g_i$ and $h_i$.
In the present investigation, they will be constrained in the next section by fitting them to
the phase shifts of pion-nucleon scattering.

Turning to the weak component of the meson--baryon Lagrangian,
the form of the lowest order 
Lagrangian is
\beq
{\cal L}_{\phi B}^{W }  =  \:
d \, \langle \bar{B}  \{ h_+ , B\} \rangle  + \:
f \, \langle \bar{B}  [ h_+ , B ] \rangle   ,
\eeq
where we have defined
\beq
h_+ = u^{\dagger} h u +  u^{\dagger} h^{\dagger} u   , 
\eeq
which transforms as a matter field,
with
$h^{a}_{b} = \delta^{a}_{2} \delta^{3}_{b}$ being
the weak transition matrix.
The weak LECs $d,f$ are the only weak counterterms considered in most previous 
calculations \cite{DGH,Bij,Jen} and
use of this Lagrangian does not provide a simultaneously 
satisfactory fit to
$s$ and $p$ waves both at the tree and the one-loop level.
As mentioned in the Introduction, a fit to the decay amplitudes has been shown to be possible
neglecting final state interactions and under the
inclusion of higher chiral orders of the weak Lagrangian due to the proliferation
of new unknown LECs \cite{BH1}. However, such a fit is not unique, and we
refrain from performing one here.
In this work, we will rather employ the weak Lagrangian at the tree level and then
append final state interactions of the meson-baryon system
within the coupled channel model which is described
in Section~\ref{sec:fsi}. This will help to understand the role of final state 
interactions and the pertinent resonances in nonleptonic hyperon decays.

\subsection{Tree level results}
\begin{figure}[htb]
\centering
\includegraphics[width=13cm,angle=0]{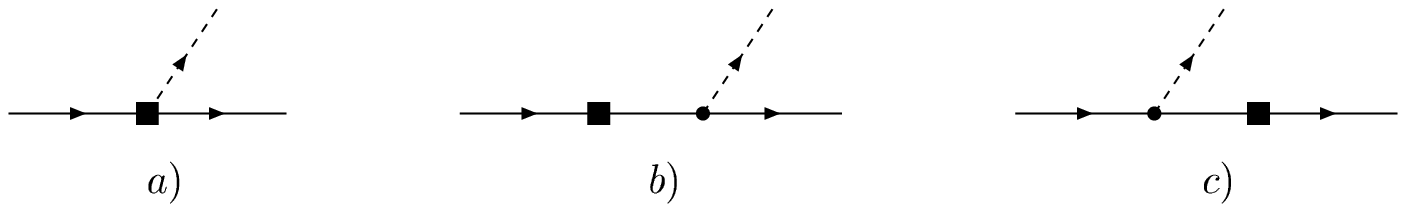}
\caption{Shown are the diagrams that contribute at the tree level. $(a)$ contributes
to the $s$ waves, whereas $(b)$ and  $(c)$ contribute to $p$ waves. Solid and dashed lines
denote baryons and pions, respectively. Solid squares and circles
are vertices of the weak and strong interactions, respectively.}
\label{fig:tree}
\end{figure}

We first calculate the tree level diagrams for the nonleptonic hyperon decays
which are depicted in Fig.~\ref{fig:tree}.

\noindent
The $s$ wave decay amplitudes $A^{(tr)}$ at the tree level are given by
\beq  \label{atree}
A_{\Sigma^+ n}^{(tr)} = 0 , \q  A_{\Sigma^- n}^{(tr)} = \frac{1}{\sqrt{2} f_\pi} (d-f) , \q
A_{\Lambda p}^{(tr)} = - \frac{1}{2\sqrt{3} f_\pi} (d+3f) , \q
A_{\Xi^- \Lambda}^{(tr)} = - \frac{1}{2\sqrt{3} f_\pi} (d-3f)  .
\eeq
For the $p$ wave amplitudes $B^{(tr)}$ one finds at the tree level
\beqa \label{btree}
B_{\Sigma^+ n}^{(tr)} &=& - \frac{M_\Sigma + M_N}{\sqrt{2} f_\pi} \left( \frac{D [d-f]}{M_\Sigma - M_N}  
     +  \frac{D [d+3f]}{3 [M_\Lambda - M_N]}  \right) , \no \\[3pt]
B_{\Sigma^- n}^{(tr)} &=& - \frac{M_\Sigma + M_N}{\sqrt{2} f_\pi} \left( \frac{F [d-f]}{M_\Sigma - M_N}  
     +  \frac{D [d+3f]}{3 [M_\Lambda - M_N]}  \right) , \no \\[3pt]
B_{\Lambda p}^{(tr)} &=&  \frac{M_\Lambda + M_N}{2 \sqrt{3} f_\pi} \left( \frac{[D+F] 
  [d+3f]}{M_\Lambda - M_N}  +  \frac{2 D [d-f]}{M_\Sigma - M_N}  \right) , \no \\[3pt]
B_{\Xi^- \Lambda}^{(tr)} &=& - \frac{M_\Xi + M_\Lambda}{2 \sqrt{3} f_\pi} \left( \frac{[D-F] 
         [d-3f]}{M_\Xi -M_\Lambda} +  \frac{2 D [d+f]}{M_\Xi -M_\Sigma}  \right) .
\eeqa
As mentioned in the Introduction, a decent fit to both $s$ and $p$ waves
in terms of the weak couplings $d$ and $f$ is not possible and we postpone the
discussion of the numerical results of such a tree level fit to Sec.~\ref{sec:results}.
The next step in our approach consists of appending final state interactions 
for the outgoing meson-baryon system.

\section{Final state interactions} \label{sec:fsi}
\begin{figure}[ht] 
\[
\parbox{2.3cm}{\centering\includegraphics[scale=0.8]{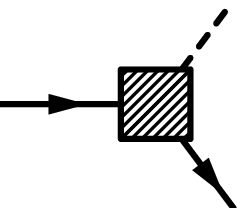}}
=
\parbox{2.4cm}{\centering\includegraphics[scale=0.8]{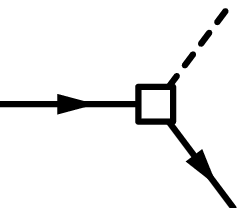}}
+
\parbox{4.0cm}{\centering\includegraphics[scale=0.8]{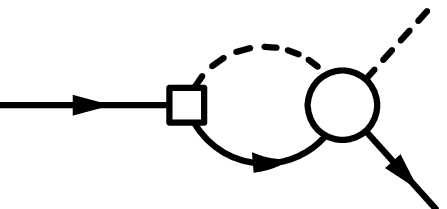}}
+
\parbox{5.0cm}{\centering\includegraphics[scale=0.8]{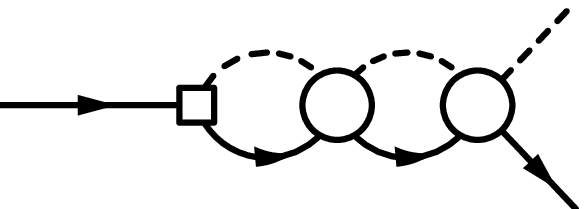}}
+\,\,\ldots
\]
\caption{The full weak decay amplitude (shaded square) is the  sum of the tree level
weak vertex (empty square) and diagrams which include strong rescattering processes
of a meson-baryon pair to all orders (empty circles).}
\label{fig:bse}
\end{figure}

\begin{figure}[ht]
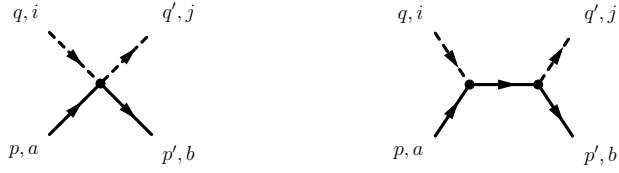

\centering 
\includegraphics[width=2.5cm]{fig3a.ps} \qquad \qquad \qquad
\includegraphics[width=3.0cm]{fig3b.ps}
\caption{Contact interaction and $s$-channel Born term
         for meson-baryon scattering.
         Solid and dashed lines denote the baryons and pseudoscalar mesons,
         respectively. $p, p', q, q'$ and $a,b,i,j$ are the four-momenta and flavor
         indices of the particles, respectively.}
\label{figmes}
\end{figure}

Final state interactions are expected to be moderate for nonleptonic
hyperon decays, and they have therefore been
omitted in previous work.
This assumption is based on Watson's theorem which states that the phase of
the amplitudes is given by the strong baryon-meson phase shifts, if $CP$ is
conserved. Aside from the $\pi N$ system, these phase shifts are not known
precisely, but are estimated to be about $10^\circ$.
The main purpose of this work is to critically re-examine this assumption by
including explicitly final state interactions within a coupled channel approach.
This is achieved by taking into account the rescattering of the meson and baryon
both in $s$ and $p$ waves as depicted in Fig.~\ref{fig:bse}. This procedure
implies that the initial baryon decays into an intermediate
meson-baryon pair which then rescatters into the final states of the decay. 
In order to describe the rescattering process, we derive the effective meson-baryon
potentials from the strong effective Lagrangian of Sec.~\ref{sec:nonlep}.
In our model the effective potentials for
the meson-baryon scattering process $B_a \phi_i \rightarrow B_b \phi_j$
are obtained by the diagrams
shown in Fig.~\ref{figmes} \cite{BMW}.

At leading order the scattering amplitude is derived 
from the effective Lagrangian ${\cal L}_{\phi B}^{(1)}$
and reads in the center-of-mass frame
\beq
V = N_a N_b \chi_b^\dagger \Big[ g(s,\vartheta) + i h(s,\vartheta) 
        ( {\bf p'} \times {\bf p}) \cdot  \mbox{\boldmath$ \sigma$} \Big]  \chi_a,
\eeq
where $\chi_i$ are Pauli-spinors and the normalization constants are given by
$N_i = \sqrt{M_i + E_i}$ with $M_i, E_i$ being the physical mass and energy
of the baryon $i$, respectively. Furthermore, $s$ is the invariant energy squared
and $\vartheta$ the center-of-mass angle between the incoming and outgoing baryon.
The functions $g$ and $h$ are given by
\beqa
g(s,\vartheta) &=& A(s) \left( 1 - \frac{{\bf p'} \cdot {\bf p}}{N_a^2 N_b^2}   \right)
  + B(s) \left( \sqrt{s} -M_a + \frac{\sqrt{s} +M_a}{N_a^2 N_b^2}{\bf p'} \cdot {\bf p} \right) \no \\
h(s,\vartheta) &=& - A(s) \frac{1}{N_a^2 N_b^2} + B(s) \frac{\sqrt{s} +M_a}{N_a^2 N_b^2}
\eeqa
with
\beqa
A(s) &=& - \frac{1}{4 f_\pi^2} (M_a -M_b) \sum_k f_{kji} f_{kab}
 + \frac{1}{2 f_\pi^2} \sum_k (D d_{jbk} + F f_{jbk} ) (D d_{iak} + F f_{iak} ) (M_k+M_b)
 \frac{s-M_a^2}{s-M_k^2} \no \\
B(s) &=& - \frac{1}{2 f_\pi^2} \sum_k f_{kji} f_{kab}
 - \frac{1}{2 f_\pi^2} \sum_k (D d_{jbk} + F f_{jbk} ) (D d_{iak} + F f_{iak} ) 
 \frac{s+M_a M_b +M_a M_k +M_k M_b  }{s-M_k^2} . \no \\
\eeqa 
We employed the physical values of the baryon masses instead of the common octet mass
$\mnod$ which is consistent at the order we are working 
and use of the physical masses is also mandatory, in order
to reproduce the correct threshold positions of the different
channels involved. The structure constants are defined by $f_{ijk} = \langle [\lambda_i,
\lambda_j] \lambda_k^\dagger \rangle$ and $d_{ijk} = \langle \{\lambda_i,
\lambda_j\} \lambda_k^\dagger \rangle$ with $\lambda_i$ being the generators of
the $SU(3)$ algebra normalized according to $\langle \lambda_i \lambda_j^\dagger\rangle = \delta_{ij} $.

The partial wave amplitude which contributes to the $s$ waves reads
\beq
V_{0+}^{(1)} = N_a N_b \Big(A(s) +B(s) [\sqrt{s} -M_a] \Big),
\eeq
while the $p$ wave contribution is
\beq
V_{1-}^{(1)} = \frac{|{\bf p}||{\bf p'}|}{N_a N_b} \Big(-A(s) +B(s) [\sqrt{s} +M_a] \Big) ,
\eeq
and the remaining partial wave amplitude $V_{1+}$ does not contribute in
nonleptonic hyperon decays
due to angular momentum conservation. The superscript denotes the chiral
order of the effective meson-baryon Lagrangian used.

Inclusion of the meson-baryon Lagrangian at second chiral order, Eq.~(\ref{bar2}),
yields the additional contributions
\beq
V_{0+}^{(2)} = N_a N_b \Big( M - N \frac{|{\bf p}|^2|{\bf p'}|^2}{3 N_a^2 N_b^2} + L
      \Big[ \frac{q_0 |{\bf p'}|^2}{N_b^2} + \frac{q'_0 |{\bf p}|^2}{N_a^2} 
      + \frac{2 |{\bf p}|^2|{\bf p'}|^2}{3 N_a^2 N_b^2} \Big]   \Big)
\eeq
and
\beq
V_{1-}^{(2)} = N_a N_b |{\bf p}| |{\bf p'}| \Big( \frac{1}{3} N - M \frac{1}{ N_a^2 N_b^2} - L
      \Big[ \frac{q_0 }{N_a^2} + \frac{q'_0 }{N_b^2} + \frac{2 }{3} \Big]   \Big)
\eeq
with
\beqa
M &=& - \frac{1}{2 f_\pi^2} b_D \left( \langle \lambda_b^\dagger\{\{\lambda_i, 
\{\lambda_j^\dagger,\chi\}\} ,\lambda_a\} \rangle
+\langle \lambda_b^\dagger\{\{\lambda_j^\dagger, 
\{\lambda_i,\chi\}\} ,\lambda_a\} \rangle\right)\no \\
&&- \frac{1}{2 f_\pi^2} b_F \Big( \langle \lambda_b^\dagger[\{\lambda_i, 
\{\lambda_j^\dagger,\chi\}\} ,\lambda_a] \rangle +\langle \lambda_b^\dagger [ \{\lambda_j^\dagger, 
\{\lambda_i,\chi\}\} ,\lambda_a ] \rangle \Big)
- \frac{2}{f_\pi^2} b_0  \delta_{ab} \langle \{\lambda_i, 
\lambda_j^\dagger\} \chi \rangle \no \\
&&+\frac{2}{f_\pi^2} (v \cdot q) (v \cdot q') \Big( d_1 \sum_k \left[ f_{j k a}
  d_{i k b} +  f_{i a k}  d_{j b k} \right]  
  + d_2 \sum_k  \left[ f_{j k a} 
  f_{i k b} +  f_{i a k}  f_{j b k} \right] \no \\
 && \qq \qq \qq 
  +d_3 \left[ \delta_{ib} \delta_{ja} +  \delta_{i^\dagger a} \delta_{j^\dagger b}\right] 
  + 2 d_4  \delta_{ab} \delta_{ij} \Big)  \no \\[5pt]
N &=&   \frac{2}{f_\pi^2} \Big( g_1 \sum_k  \left[ f_{j k a} 
  d_{i k b} +  f_{i a k}  d_{j b k} \right]  
  + g_2 \sum_k  \left[ f_{j k a} 
  f_{i k b} +  f_{i a k}  f_{j b k} \right] \no \\
 && \qq 
  +g_3 \left[ \delta_{ib} \delta_{ja} +  \delta_{i^\dagger a} \delta_{j^\dagger b}\right] 
  + 2 g_4  \delta_{ab} \delta_{ij} \Big)  \no \\[5pt]
L &=&   \frac{2}{f_\pi^2} \Big( 2 h_1 \sum_k  f_{k j i} d_{kab}
   + 2 h_2 \sum_k  f_{k j i} f_{kab} 
   + h_3 \left[ \delta_{ib} \delta_{ja} -  \delta_{i^\dagger a} \delta_{j^\dagger b}\right] \Big),
\eeqa
where we have introduced the notation 
$\langle  \lambda_i \lambda_j
\rangle  = \delta_{i^\dagger j}$.

After deriving the effective potentials for meson-baryon scattering both
for $s$ and $p$ waves, they are iterated in a BSE which couples the different
physical channels. The following channels contribute via final state interactions to nonleptonic
hyperon decays: $\pi N$, $\eta N$,
$K \Lambda$, $K \Sigma$ for $S=0$, $I=1/2$ , $\pi N$, $K \Sigma $ for $S=0$, $I=3/2$,
and $\pi \Lambda$, $\pi \Sigma$, $\bar{K} N$,
$\eta \Sigma$, $K \Xi$ for $S=-1$, $I=1$.
In the Bethe-Salpeter formalism the total strong $T$ matrix can be written
in channel space in the matrix form $T = (1 + V \cdot G)^{-1} V$ \cite{OM, BMW}.
In a similar fashion, the total weak decay amplitude is given by
\beqa \label{eq:coupled}
A &=& [ 1 + (V_{0+}^{(1)}+V_{0+}^{(2)}) \cdot G ]^{-1} A^{(tr)} \no \\
B &=& [ 1 + (V_{1-}^{(1)}+V_{1-}^{(2)}) \cdot G ]^{-1} B^{(tr)} ,
\eeqa
where $G$ is the finite part of the scalar loop integral $\tilde{G}$
\begin{equation}
\tilde{G}(q^2) = \int \frac{d^d l}{(2 \pi)^d} 
\frac{i}{[ (q-l)^2 - M_B^2 + i \epsilon]
   [ l^2 - m_\phi^2 + i \epsilon] } 
\end{equation}
and is understood to be a diagonal matrix in channel space.
One obtains, {\it e.g.}, in dimensional regularization
\begin{eqnarray}
G(q^2) &=& \frac{1}{32 \pi^2 q^2} \Bigg\{ q^2
\left[ \ln\Big(\frac{m_\phi^2 }{\mu^2}\Big) +
\ln\Big(\frac{M_B^2 }{\mu^2}\Big) -2 \right] 
+ (m_\phi^2 - M_B^2)  \ln\left(\frac{m_\phi^2 }{M_B^2}\right) \no \\
&&  - 8 \sqrt{q^2}|\mbox{\boldmath$q$}_{cm}| \; \mbox{artanh }
\left(\frac{2 \sqrt{q^2}|\mbox{\boldmath$q$}_{\scriptstyle{cm}}|}{
(m_\phi + M_B )^2 - q^2} \right) \Bigg\} ,
\end{eqnarray}
where $\mu$ is the regularization scale. 
Note that the expression for $G$ is regularization scale dependent and $\mu$ is treated as a
so-called finite range parameter \cite{KSW, KWW, BMW}. Along with the unknown coupling
constants of the strong effective Lagrangian of Eq.~(\ref{bar2})
it is used in this approach to reproduce the
experimental phase shifts of pion-nucleon scattering at the relevant energies.
In Eq.~(\ref{eq:coupled}) the $s$ and $p$ wave amplitudes have been summarized
in column vectors $A^{(tr)}$ and $B^{(tr)}$.

The imaginary part of the matrix $T^{-1}$ of the strong interactions is identical with the imaginary
piece of the fundamental scalar loop integral $\tilde{G}$
above threshold,
\begin{equation} \label{unit}
\mbox{Im} T^{-1} = - \frac{|\mbox{\boldmath$q$}_{cm}|}{8 \pi \sqrt{s}} 
\end{equation}
with
$\mbox{\boldmath$q$}_{cm}$ being the three-momentum in the
center-of-mass frame of the channel under consideration. 
Eq.~(\ref{unit}) is the restriction which unitarity imposes
on the $T$ matrix for each partial wave.
The BSE (\ref{eq:coupled}) amounts to a summation of a bubble chain
as depicted in Fig.~\ref{fig:bse} and
isospin invariance of the strong interactions guarantees that the solutions
of the BSE in (\ref{eq:coupled}) satisfy the isospin relations in 
Eq.~(\ref{iso}).

\section{Numerical Results} \label{sec:results}

\begin{figure}[t]
\centering
\begin{picture}(300,410)
\put(0,310){\makebox(100,120){\epsfig{file=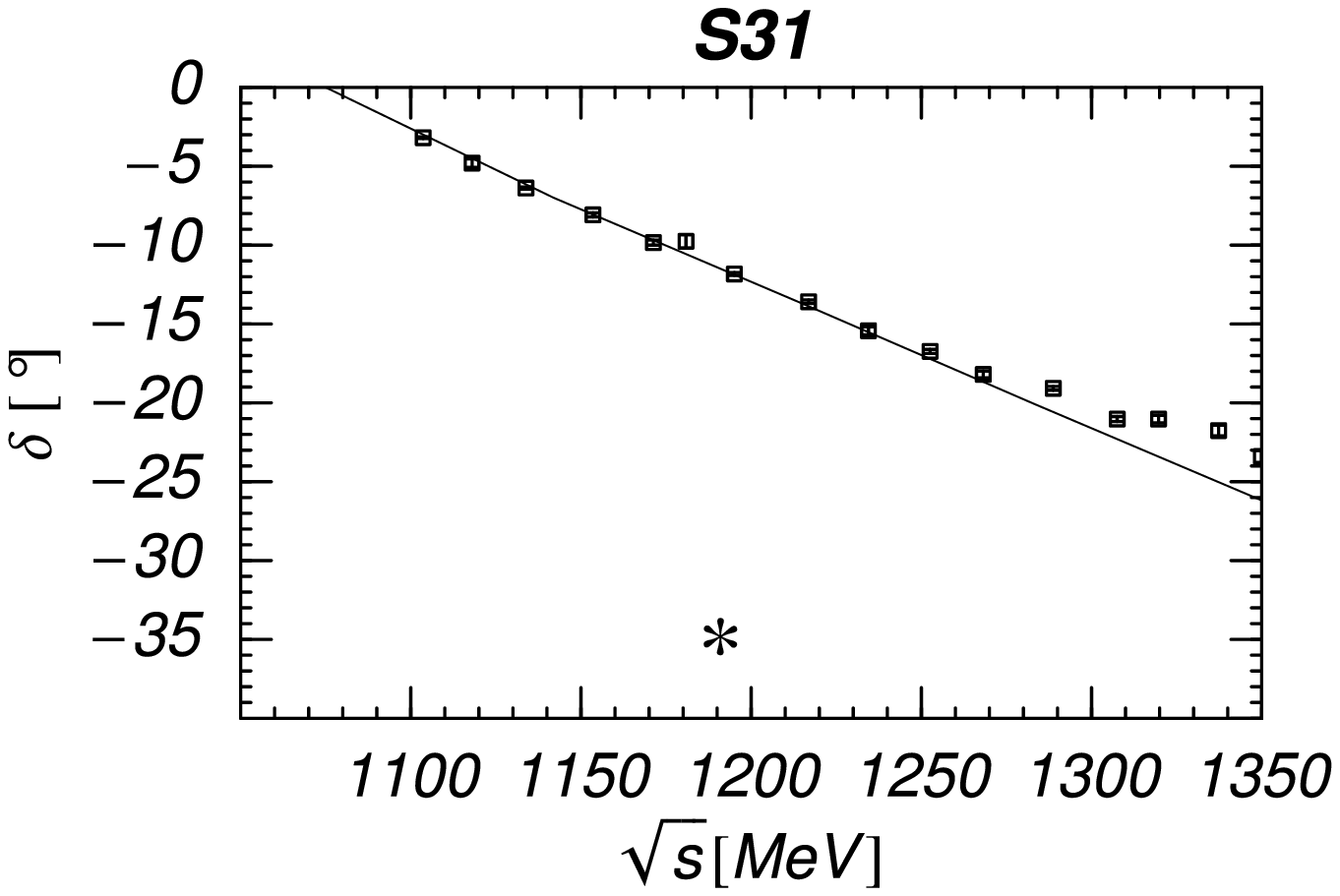,width=6.5cm,angle=0}}}
\put(200,310){\makebox(100,120){\epsfig{file=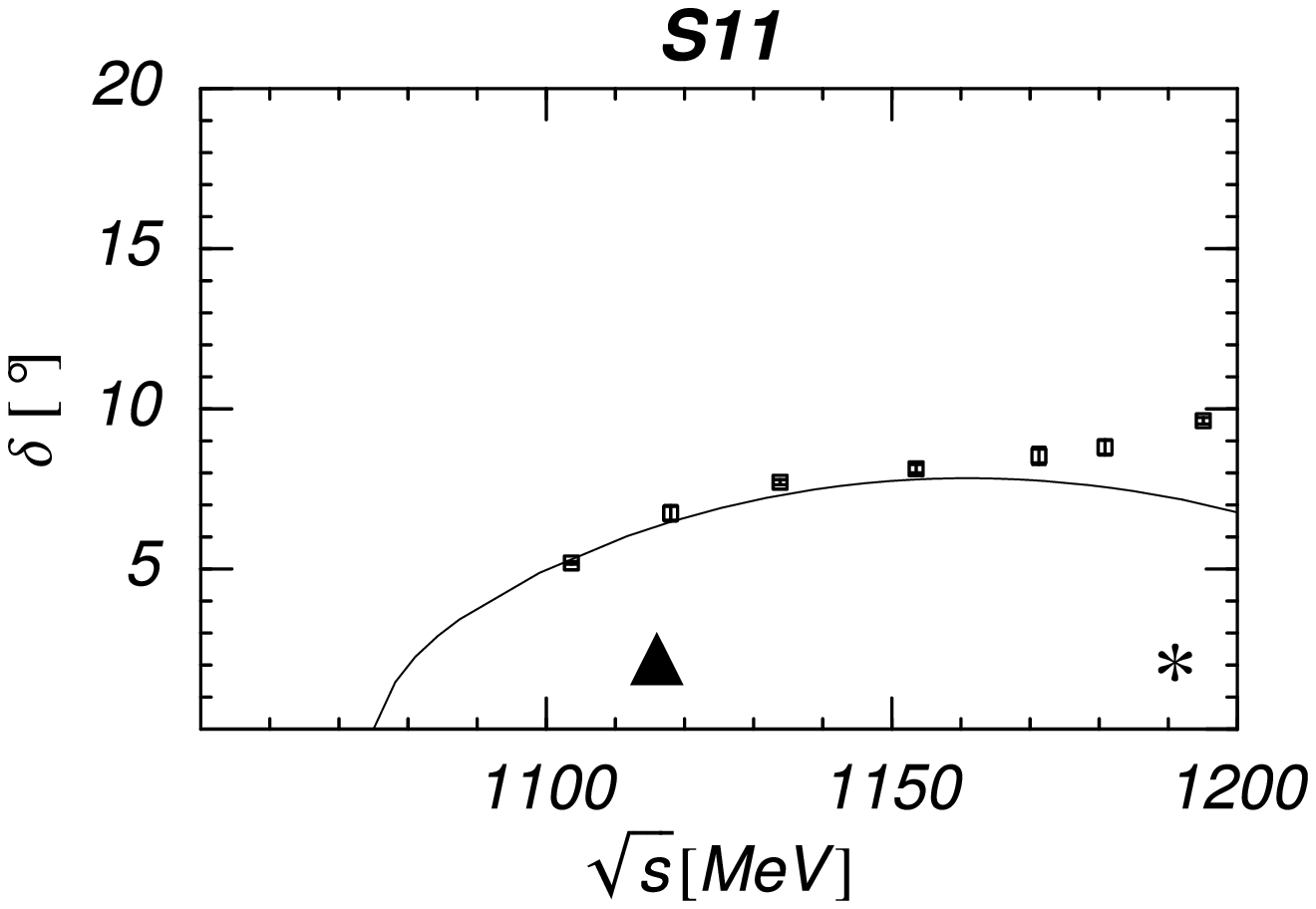,width=6.5cm,angle=0}}}
\put(0,165){\makebox(100,120){\epsfig{file=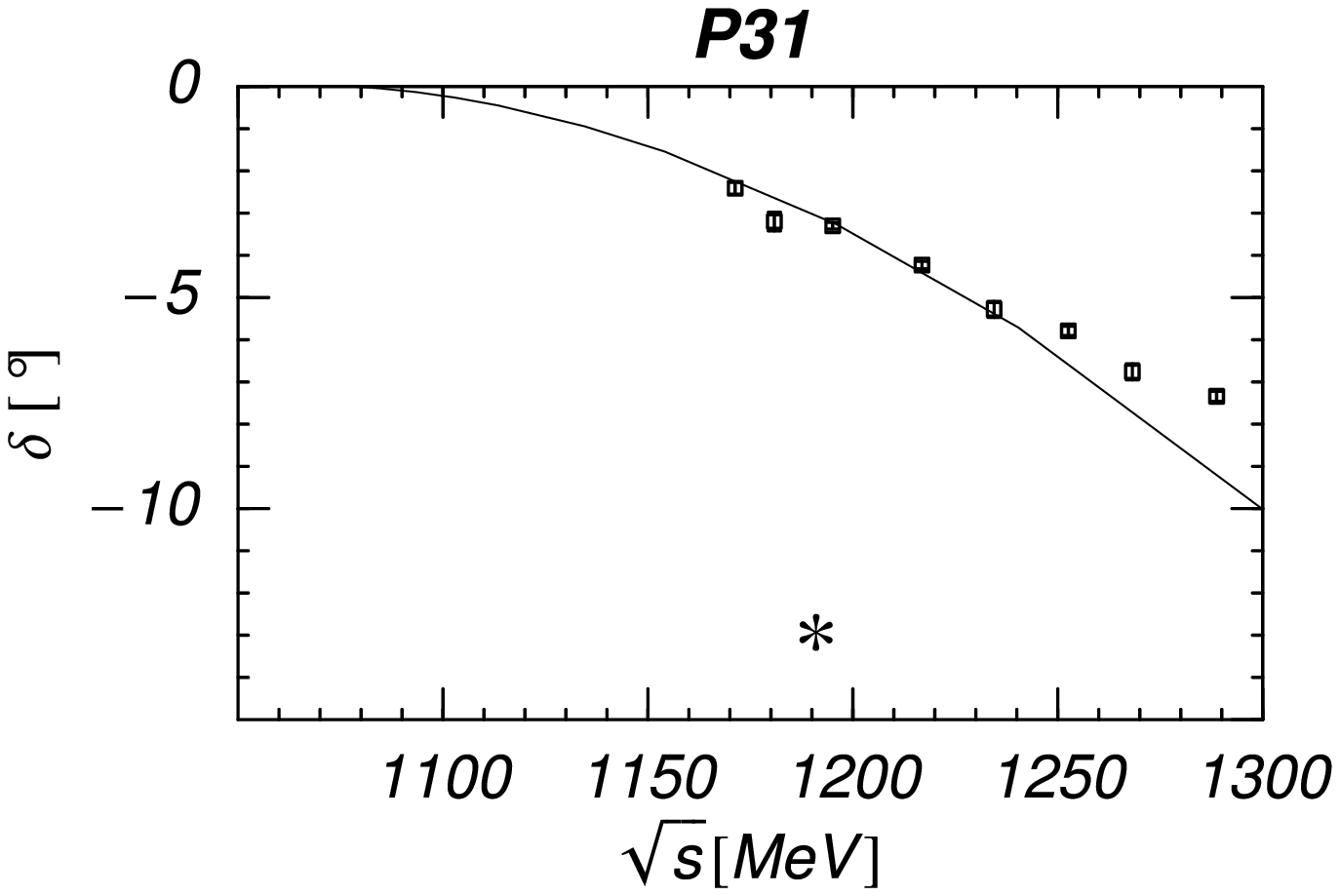,width=6.5cm,angle=0}}}
\put(200,165){\makebox(100,120){\epsfig{file=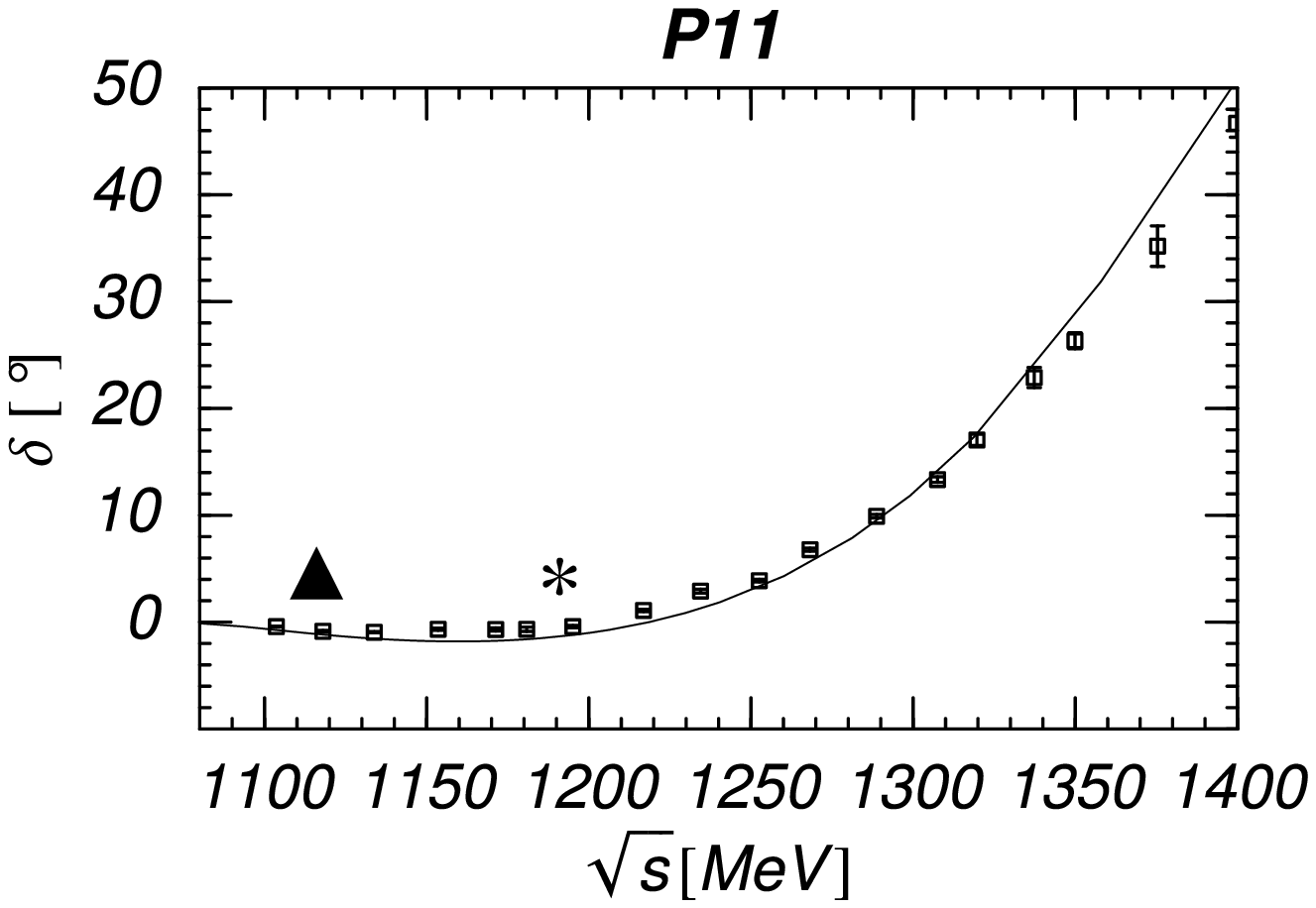,width=6.5cm,angle=0}}}
\put(0,20){\makebox(100,120){\epsfig{file=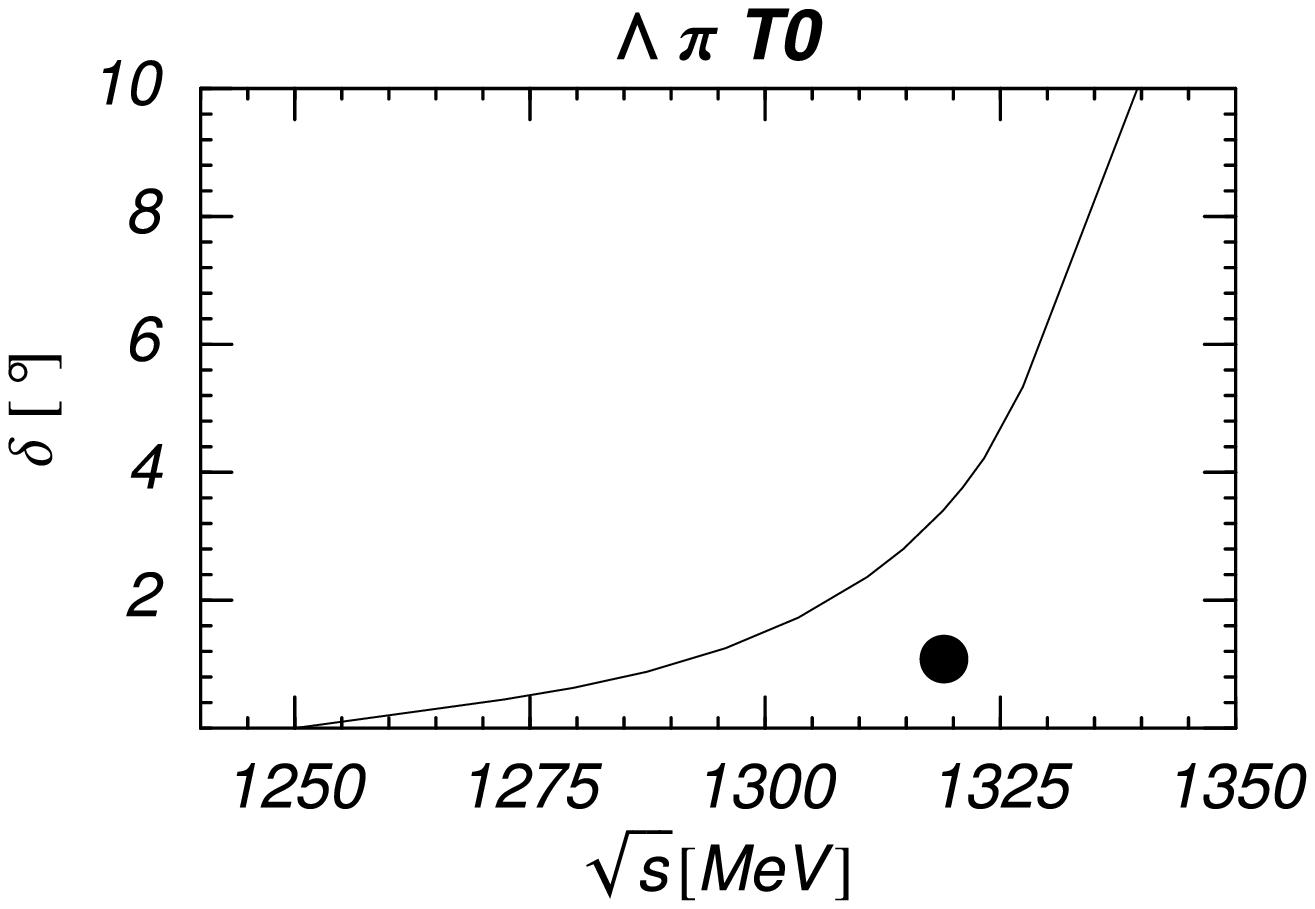,width=6.5cm,angle=0}}}
\put(200,20){\makebox(100,120){\epsfig{file=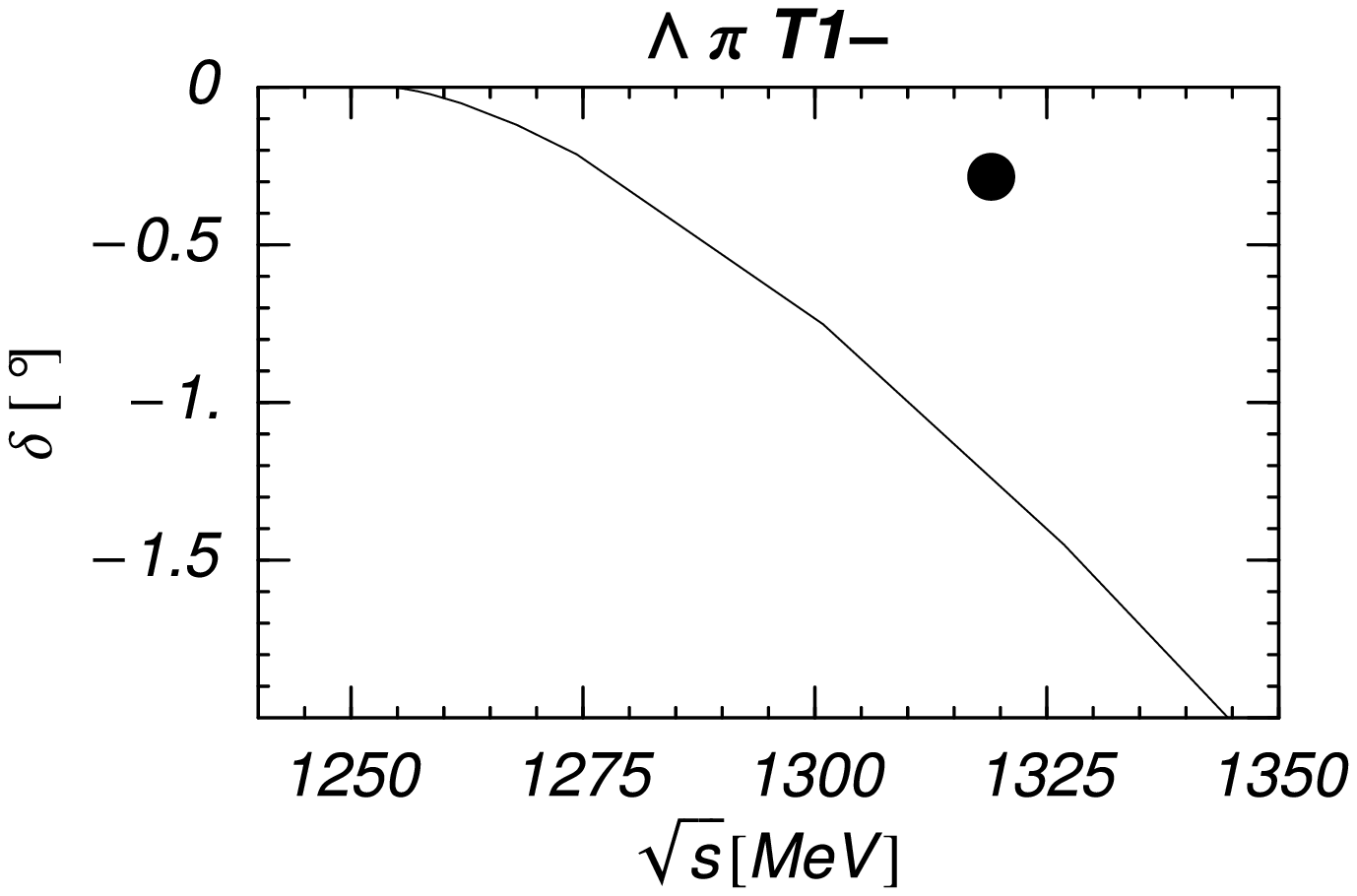,width=6.5cm,angle=0}}}
\put(100,395){\large $a)$}
\put(300,395){\large $b)$}
\put(110,250){\large $c)$}
\put(280,250){\large $d)$}
\put(6,105){\large $e)$}
\put(310,105){\large $f)$}
\end{picture}
\caption{Shown are fits to the phase shifts of $\pi$-$ N$ (diagrams $a$ to $d$)
and $\pi$-$\Lambda$ ($e$, $f$) partial wave amplitudes.
The masses of the $\Lambda, \Sigma, \Xi$ are denoted by a triangle, asterisk, and circle, 
respectively. }
\label{fig:phase}
\end{figure}
In this section, we compare the numerical results of our Bethe-Salpeter approach
in Eq.~(\ref{eq:coupled}) for the observables $\Gamma, \alpha$ and $\phi$
with the experimental data, while values for $\beta$ can be deduced directly
by employing Eq.~(\ref{observ}).
The unknown coupling coefficients of the strong effective Lagrangian are
constrained by reproducing the phase shifts of pion-nucleon partial wave
amplitudes.
The results are shown in Fig.~\ref{fig:phase}.
We are clearly able to obtain a good fit to the phase shifts in the energy
range we are working in. The fit also yields predictions for $\pi$-$\Lambda$
scattering for which---to our knowledge---experimental data is not available.
Our results for the $s$ wave  $\pi$-$\Lambda$ phase shifts are slightly above
those of \cite{OM2}. However, the authors of \cite{OM2} made use of the leading
meson-baryon Lagrangian only and omission of the contact terms from the
Lagrangian of second chiral order reduces our result for this phase shift,
bringing it to better agreement with \cite{OM2}.

For the fit presented in Fig.~\ref{fig:phase} we employed the values
(all in units of GeV$^{-1}$) $b_D=0.066$, $b_F=-0.185$, $b_0 = -0.254$,
$d_1=-0.6$, $d_2=0.14$,
$d_3 = 0$, $d_4=-0.37$, $g_1=0.5$, $g_2=0.5$, $g_3=0$, $g_4=0.2$,
$h_1=0.25$, $h_2 = 0.25$, $h_3=0$
and the regularization scale $\mu =1.1$ GeV in all channels.
The fit is, of course, not unique, since small changes in some of the chiral
parameters yield equally good representations of the phase shifts.
Nevertheless, it is sufficient to work with a particular choice of parameters
which ensures the accurate inclusion of the final state interactions.
After constraining the coefficients of the strong effective Lagrangian
we are left with the two weak parameters $d$ and $f$. Performing a fit for 
$d$ and $f$ to the decay observables $\Gamma, \alpha$ and $\phi$ of the seven nonleptonic hyperon decays
yields the values $d=0.31 \times 10^{-7} $ GeV and $f=-0.38\times 10^{-7}$ GeV.%
\begin{table} 
\begin{center}
\begin{tabular}{ccccccc}
	&$\Gamma_{\mbox{exp}}$ [$\mu$eV] & $\alpha_{\mbox{exp}}$&
$\phi_{\mbox{exp}} [^\circ]$ &
$\Gamma$ [$\mu$eV] & $\alpha$&$\phi [^\circ]$\\ 
\hline
\\
$\Lambda \rightarrow p \pi^-$	& $1.597 \pm 0.012$ &
$0.642 \pm 0.013 $&$ -6.5 \pm 3.5$ & 1.70 &0.28& $-1.0$\\
$\Lambda \rightarrow n \pi^0$	& $0.895 \pm 0.007$ &
$0.65 \pm 0.05$ & --- &	0.85 & 0.28 &$-1.0$\\
$\Sigma^+ \rightarrow n \pi^+$	& $3.966 \pm 0.013$ &
$0.068 \pm 0.013$ &$ 167 \pm 20$& 0.82 & $-0.86$ & 59.2\\
$\Sigma^+ \rightarrow p \pi^0$	& $4.233 \pm 0.014$&
$-0.980 \pm 0.016$ &$ 36 \pm 34$ &2.55 & $-0.60$ & $-1.4$\\
$\Sigma^- \rightarrow n \pi^-$	& $4.450 \pm 0.033$ &
$-0.068 \pm 0.008$ &$ 10 \pm 15$ &2.83 & 0.0 & 0.22\\
$\Xi^0 \rightarrow \Lambda \pi^0$ & $2.270 \pm 0.070$ &
$-0.411 \pm 0.022$ &$ 21 \pm 12$ &2.30 & $-0.27$ & 1.76\\
$\Xi^- \rightarrow \Lambda \pi^-$ & $4.016 \pm 0.037$ &
$-0.456 \pm 0.014$ &$ 4 \pm 4$ &4.61 & $-0.27$ & 1.76\\
\end{tabular}
\end{center}
\caption{Experimental values \cite{pdg} and results of our fit including
final state interactions for all decay channels.}
\label{tab:fit}
\end{table}
The results for this fit are shown in Table~\ref{tab:fit} and compared with
the experimental values. 
Despite the inclusion of final state interactions,
we are clearly not able to obtain a good fit for the decays, in particular
for $\Sigma^+ \rightarrow n \pi^+$.
These numbers must be compared with the results from a pure tree
level calculation, Eqs.~(\ref{atree}) and (\ref{btree}), which
are summarized in Tab.~\ref{tab:tree} for two different choices
of $d$ and $f$.
We first employ $d=0.17\times 10^{-7}$ GeV  and $f= -0.4 \times 10^{-7} $ GeV
which are obtained from a tree level fit
to the $s$ waves omitting final state interactions.
Second, we use the values 
$d=0.31 \times 10^{-7} $ GeV and $f=-0.38\times 10^{-7}$ GeV from
the above mentioned fit of the BSE approach. Both results are given in
Tab.~\ref{tab:tree} and
comparison with Tab.~\ref{tab:fit} shows that the inclusion of
final state interactions provides an improved overall fit, {\it e.g.},
it yields the correct sign for $\alpha$ in $\Xi$ decays which is not the case
for the tree level fit.  We observe that final state 
interactions have sizeable effects for nonleptonic hyperon decays,
{\it i.e.} much larger than the anticipated 10\% level , if the same
values for $d$ and $f$ are employed.
Furthermore, the inclusion of final state interactions allows for a
prediction of
the decay parameter $\phi$ which vanishes in the tree level approximation.
\begin{table} 
\begin{center}
\begin{tabular}{ccccc}
	&$\Gamma_1$ [$\mu$eV] & $\alpha_1$&
$\Gamma_2$ [$\mu$eV] & $\alpha_2$\\ 
\hline
\\
$\Lambda \rightarrow p \pi^-$	& 1.66 &
0.69 & 0.97 &0.39\\
$\Lambda \rightarrow n \pi^0$	& 0.83 &
0.69 & 0.48 & 0.39\\
$\Sigma^+ \rightarrow n \pi^+$	& 0.04 &
0& 0.56 & 0 \\
$\Sigma^+ \rightarrow p \pi^0$	& 2.46 &
$-0.53$ & 3.6 &$-0.53$  \\
$\Sigma^- \rightarrow n \pi^-$	& 4.71 &
$-0.36$ & 6.67  &0.01  \\
$\Xi^0 \rightarrow \Lambda \pi^0$ & 1.70 &
0.13 & 1.95  &$-0.31$ \\
$\Xi^- \rightarrow \Lambda \pi^-$ & 3.41 &
0.13  & 3.90  &$-0.32$  \\
\end{tabular}
\end{center}
\caption{Given are the tree level contributions using two sets of parameters.
1:  $d=0.17\times 10^{-7}$ GeV and $f= -0.4 \times 10^{-7} $ GeV
from a tree level fit to the $s$ waves.
2: Results using the values of
$d$ and $f$ from our fit including final state interactions, $d=0.31 
\times 10^{-7} $ GeV and $f=-0.38\times 10^{-7}$ GeV.
There is no contribution to $\phi$ at the tree level.}
\label{tab:tree}
\end{table}

\section{Conclusions} \label{sec:con}
In this work, we have investigated the importance of final state interactions
in nonleptonic hyperon decays within a coupled channel approach. First,
the tree level contributions to the decays have been calculated from the 
lowest order weak Lagrangian. In a second step, we consider final state interactions 
of the decay particles which are modelled in this approach by calculating
the effective meson-baryon scattering potentials and iterating them in a Bethe-Salpeter
equation. The effective potentials are obtained from the strong effective
Lagrangian at next-to-leading order. The inclusion of the strong counterterms of
second chiral order is necessary, in order to obtain better agreement with
experimental phase shifts from meson-baryon scattering and to ensure a proper
treatment of the final state interactions. 
In fact, we are able to reproduce accurately the 
experimental phase shifts of pion-nucleon scattering in the energy 
range we are working in.

The iteration of the effective potential in the BSE produces dynamically the 
pertinent low lying baryon resonances. Thus without including the resonances explicitly,
we can independently check their importance in nonleptonic hyperon decays.
This allows for a critical re-examination of previous work in which resonant states
have been taken into account explicitly. 
In this work, we were able to show that the generation of resonances by means
of final state interactions and omission of higher order weak LECs does not
enable a reasonable fit to nonleptonic hyperon decays in terms of just the lowest
order weak parameters, $d$ and $f$, although it decreases the discrepancy with experiment.
In \cite{BH2}, on the other hand, the resonances have been integrated out
producing weak counterterms of higher chiral orders which were parameterized
by the weak decay parameters of the resonances and a fit to the decay amplitudes
was possible.
This indicates that in \cite{BH2} the weak couplings which were obtained
from a fit to the decay amplitudes may have been overestimated. But it also
underlines the importance of higher order weak counterterms in order to obtain
a reasonable fit to both $s$ and $p$ waves.
This observation is reinforced by the investigation in the quark model \cite{LeY},
since contributions from higher resonances are included in this approach, whereas they
have been omitted both in \cite{BH2} and in the present study. In ChPT
the effects of higher resonances are hidden in contributions to higher
order coupling coefficients of the weak Lagrangian.
Furthermore, we have shown that final state interactions yield sizeable effects
in nonleptonic hyperon decays, definitely larger than the expected 10\% level,
and should not be omitted. 

We can therefore conclude that a reasonable fit to the nonleptonic hyperon decays
is not possible in ChPT without the inclusion of higher order weak counterterms.
In the BSE approach contributions from low lying resonances are sizeable and 
yield a slight improvement, but are definitely
not capable of accounting for the discrepancy with experiment .


\end{document}